\documentstyle[aas2pp4]{article}

\slugcomment{Accepted for publication in ApJ Letters}
\lefthead{Burke et al.}
\righthead{The XLF of distant X-ray Galaxy Clusters}

\def\vmax{\mbox{$V_{\rm max}$}}
\def\li{\mbox{$L_{i}$}}

\def\persec{\mbox{s$^{-1}$}}
\def\lumin{\mbox{erg \persec}}

\def\kunits{\mbox{$10^{-7}$ Mpc$^{-3}$ (10$^{44}$ \lumin)$^{\alpha-1}$}}

\def\arbest{2.15}  
\def\arsigma{0.23}
\def\krbest{2.72}  
\def\krsigma{0.12}

\def\aebest{2.22}  
\def\aesigma{0.25}
\def\kebest{4.99}  
\def\kesigma{0.77}

\def\srabest{1.77}     
\def\srasigma{0.30}
\def\srkbest{3.64}
\def\srksigma{0.16}
\def\seabest{1.78}     
\def\seasigma{0.30}
\def\sekbest{5.68}
\def\seksigma{1.04}

\begin{document}

\title{The Southern SHARC Survey: the ${\mathbf z = 0.3 - 0.7}$
Cluster XLF\footnote{Based partly on data collected at the 
European Southern Observatory, La Silla, Chile, 
and the Anglo-Australian Telescope, Siding Spring, Australia}}

\author{
D.J. Burke\altaffilmark{2,3},
C.A. Collins\altaffilmark{3},
R.M. Sharples\altaffilmark{2},
A.K. Romer\altaffilmark{4,6},
B.P. Holden\altaffilmark{5},
R.C. Nichol\altaffilmark{6}
}

\altaffiltext{2}{Department of Physics, University of Durham, South Road,
Durham DH1 3LE, UK}
\altaffiltext{3}{Astrophysics Research Institute, School of Engineering, Liverpool John 
Moores University, Byrom Street, Liverpool L3 3AF, UK}
\altaffiltext{4}{Department of Physics and Astronomy, Northwestern University,
2145 Sheridan Road, Evanston, IL 60208}
\altaffiltext{5}{Department of Astronomy and Astrophysics, University of Chicago, 
5640 S. Ellis Rd., Chicago, IL 60637}
\altaffiltext{6}{Department of Physics, Carnegie Mellon University,
5000 Forbes Avenue, Pittsburgh, PA 15213-3890}

\begin{abstract}
We present the $z = 0.3 - 0.7$
cluster X-ray luminosity function (XLF) determined
from the Southern SHARC (Serendipitous High-redshift Archival ROSAT Cluster)
survey.
Over the luminosity range
$L \sim (0.3 - 3) \times 10^{44}$ erg s$^{-1}$ (0.5 - 2.0 keV) 
the XLF is in close agreement with that of the low
redshift X-ray cluster population.
This result greatly strengthens our previous claim of no
evolution of the cluster population, at these
luminosities, at a median redshift of $z =0.44$.

\end{abstract}


\section{Introduction}

Xray clusters of galaxies are efficient tracers of the mass in 
the Universe and can be studied out to large redshifts.
Prior to the launch of the ROSAT X-ray telescope, the
only high redshift ($z > 0.3$) X-ray selected cluster sample
was that of the EINSTEIN Extended Medium Sensitivity Survey
(EMSS; Henry et al. \markcite{h92}1992;
Gioia \& Luppino \markcite{gl-94}1994).
Henry et al. \markcite{h92}(1992) used the EMSS to show 
X-ray clusters evolved `negatively' --- the space
density of high luminosity clusters
being lower in the redshift 
range $z = 0.30 - 0.60$ compared to $z = 0.14 - 0.20$.
This result was in conflict with popular models of cluster
formation (e.g. Kaiser \markcite{k86}1986) and prompted 
further significant theoretical work (e.g. Kaiser \markcite{k91}1991;
Evrard \& Henry \markcite{eh91}1991).

The maturing of the ROSAT database has sparked much recent interest in 
testing this result.
A large cluster sample with a median depth of $z \sim 0.1$, 
created from the ROSAT All Sky Survey (RASS), 
shows no sign of evolution out to $z = 0.3$ 
(Ebeling et al. \markcite{e97}1997).
Castander et al. \markcite{RIXOS}(1995) presented the first look at the
high redshift cluster population with ROSAT, 
claiming that the evolution seen by Henry et al. \markcite{h92}(1992) 
extends to luminosities $\sim 10^{44}$ erg s$^{-1}$.
We have recently shown (Collins et al. \markcite{c97}1997),
using the Southern SHARC\footnote{Serendipitous High-redshift Archival ROSAT Cluster}
sample of serendipitously detected
clusters from deep ROSAT PSPC pointings,
that the number of high redshift clusters is consistent
with a no evolution model,
in direct contrast to Castander et al. \markcite{RIXOS}(1995). 
Finally, the EMSS sample has been re-analysed in the light of new
optical and X-ray data, which indicates that the evidence for
evolution seen by Henry et al. \markcite{h92}(1992) 
is not statistically significant (Nichol et al. \markcite{n97}1997).

In this letter we present the high redshift
X-ray luminosity function (XLF)
of the Collins et al. \markcite{c97}(1997) cluster sample. 
In section \ref{xlf} we describe the calculation of the
XLF from this sample and in section \ref{discussion}
we discuss the results.
Throughout this letter we have assumed
H$_0 = 50$ km/s/Mpc and $q_0 = 0.5$ and 
quoted luminosities in the 0.5 to 2.0 keV pass band,
unless explicitly stated otherwise.

\section{Determination of the XLF}
\label{xlf}

The cluster sample used in this letter is the Southern SHARC
survey, consisting of 16 clusters in the redshift range $z = 0.3 - 0.7$
with a median redshift of $z = 0.44$,
detected in a serendipitous survey of 66 deep ROSAT PSPC 
pointings, covering a total search area of $17.7$ deg$^{2}$. This area is 
slightly larger than that used in Collins et al. \markcite{c97}(1997) due
to the use of a different central source mask. Each pointing satisfies
the following criteria: exposure time greater than 10 ks, 
$|b| > 20^{\circ}$, and $\delta < 20^{\circ}$. Further details of the 
sample are given in Collins et al. \markcite{c97}(1997)
and a description of the full survey will be presented in a
future paper.

For each cluster the count rate was measured from the background-subtracted 
count rate image in the 0.5 to 2.0 keV pass band. 
The chosen aperture encloses 80\% of
the light and was determined from a convolution of a King surface brightness 
profile with a model of 
the off-axis point spread function, using a cluster core radius of 250 kpc 
and $\beta=2/3$ (Jones \& Forman \markcite{j84}1984). 
Justification for the adopted values of these cluster parameters is given below.
After accounting for the flux falling outside this aperture,
we used a thermal bremsstrahlung spectrum of temperature 6 keV (typical
of nearby clusters) 
and a Galactic absorption model to convert the total count rate
into a flux, and then into rest frame luminosities, $L$,
in both the 0.5 to 2.0 keV and 0.3 to 3.5 keV pass bands.
We chose to use a thermal bremsstrahlung model because
cluster X-ray emission is dominated by the 
continuum radiation rather than line emission
(e.g. Sarazin \markcite{sarazin}1988)
and for its computational ease of use. To investigate the uncertainty
introduced by using a single temperature we have used a luminosity-temperature 
relation (Wang \& Stocke \markcite{ws93}1993) to obtain iterated estimates
of the cluster luminosities. These values are 
within 1\% (0.5 - 2.0 keV) and 4\% (0.3 - 3.5 keV) of those 
obtained above and the difference is negligible compared to the Poission 
errors of the photon counts, which has a median value of $10\%$.

We have performed extensive simulations to model the effect of using an
extent criterion to select cluster candidates.
The ability to detect a given cluster depends on properties both intrinsic 
and extrinsic to the cluster ---
the relevant cluster characteristics are its luminosity, 
surface brightness profile and redshift; those for the survey are
exposure time, background count rate and off-axis angle.
The simulations assume a universal cluster X-ray surface brightness profile,
namely the King model with a core radius
of 250 kpc and $\beta = 2/3$ (Jones \& Forman \markcite{jf84}1984), 
that is independent of redshift.
Ideally one would use a distribution encompassing the true range of
cluster profiles, 
however the sensitivity and spatial resolution of the current generation of 
X-ray instruments limit the knowledge of such a distribution to
low redshift.
Our choice of profile is consistent with the average properties
of low redshift cluster samples (e.g. Henry et al. \markcite{h92}1992;
Jones \& Forman \markcite{jf84}1984).
As the simulations are computationally intensive we limited our
analysis to a binned representation of the survey
with regard to exposure time, background count rate and
off-axis angle. 
The bin sizes were chosen so as to be small enough that the
selection function did not vary significantly across
a bin whilst being large enough that the simulations could
be performed in a reasonable time.
For each bin a range of cluster luminosities ($0.1-3.0 \times 10^{44}$ \lumin) and 
redshifts ($0.2-0.9$) were used;
simulated clusters were generated by randomly distributing source
photons using the King profile convolved with a model PSPC PSF
and background photons using a flat background model.
The number of source and background photons were Poisson
distributed about the expected number.

\begin{figure}[ht]
\plotfiddle{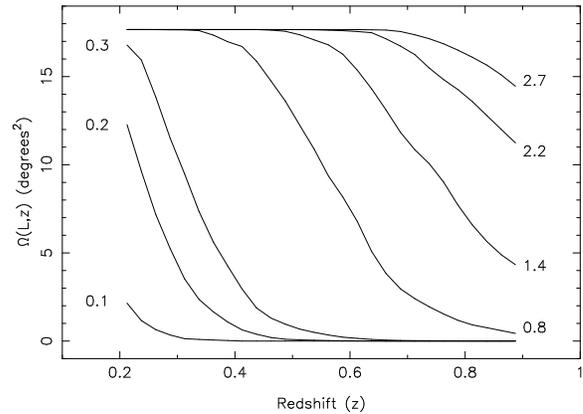}{2in}{-90}{32}{32}{-120}{180}
\caption[fig1.eps]{
The available survey area, as a function of redshift, for the Southern
SHARC survey.
Each curve corresponds to a different cluster luminosity,
labelled in units of $10^{44}$ \lumin, and was
calculated using the simulations described in the text.
\label{fig:area-vs-redshift}}
\end{figure}

\begin{figure}[ht]
\plotfiddle{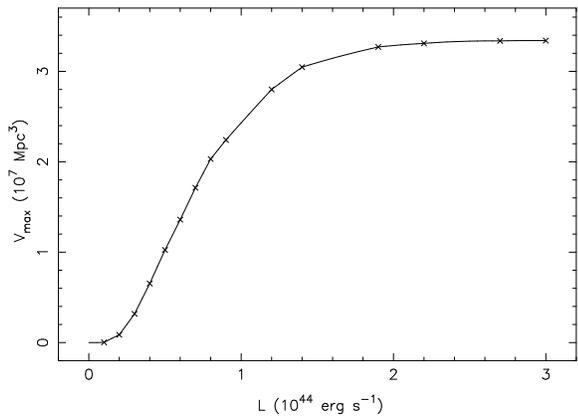}{2in}{-90}{32}{32}{-120}{180}
\caption[fig2.eps]{
The symbols indicate the values of $\vmax(L)$,
for the redshift shell $z = 0.3 - 0.7$,
calculated using equation (\protect\ref{eqn:vmax})
and the curves shown in Figure \protect\ref{fig:area-vs-redshift}.
Luminosities are in units of $10^{44}$ \lumin\ and
the solid curve is the fit to these points.
\label{fig:vol-vs-lum}}
\end{figure}

Multiplying the selection function of a given bin by the 
survey area for the bin and summing over all the bins
produces $\Omega(L,z)$,
the available survey area as a function of cluster luminosity
and redshift, shown in Figure \ref{fig:area-vs-redshift}.
The survey volume for a cluster with luminosity $L$,
in a redshift shell $z = z_1$ to $z_2$, is
$\vmax(L)$, defined as
\begin{equation}
\label{eqn:vmax}
\vmax(L) = \int_{z_1}^{z_2} \Omega(L,z) \, dV(z) \, dz,
\end{equation}
where $dV(z)$ is the volume per square degree at a redshift $z$.
Figure \ref{fig:vol-vs-lum} shows equation (\ref{eqn:vmax}) evaluated
for the redshift shell $z = 0.3 - 0.7$ using the curves from 
Figure \ref{fig:area-vs-redshift} where
the solid line indicates the fit used to interpolate these points.

We have used both parametric and non-parametric forms to define
the cluster XLF; a brief description follows and
a full account of the fitting procedure can be found in Nichol et al. 
\markcite{n97}(1997). 
The non-parametric XLF, $n(L)$, was calculated using
\begin{equation}
\label{eqn:xlf-binned}
n(L) = \sum_{i} \frac{1}{\vmax(\li) \, \Delta L},
\end{equation}
where the sum is performed over all clusters 
whose luminosities, $\li$, are within $\pm \Delta L/2$ of $L$.
For a model XLF $\phi(L)$,
the parametric form was determined by maximising
the likelihood ${\cal L}$, given by
\begin{equation}
\label{eqn:xlf-likelihood}
{\cal L} = \prod_{i}
\frac{\int_{0}^{\infty} \vmax(L) \, \phi(L) \, E_i(L) \, dL}
{\int_{0}^{\infty} \vmax(L) \, \phi(L) \, dL},
\end{equation}
where the product is over the cluster detections
and $E_i(L)$ is a normal distribution, with mean \li\ and 
standard deviation equal to the error on \li,
which models the effect of 
luminosity errors on the fitting procedure.

\begin{figure}[t]
\plotfiddle{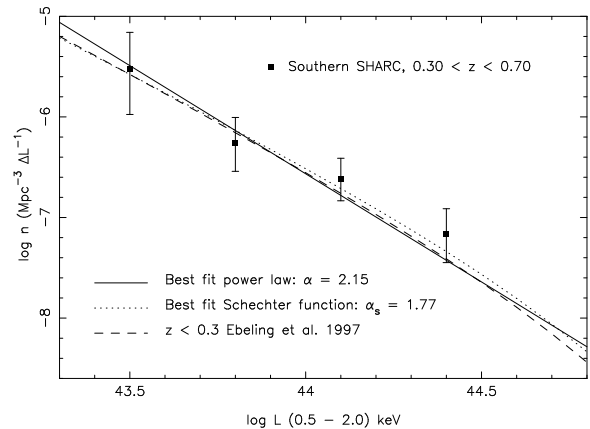}{2in}{-90}{32}{32}{-120}{180}
\caption[fig3.eps]{
The XLF of the Southern SHARC in the 0.5 to 2.0 keV pass band.
The solid line is our best power-law fit,
the dotted line our best fit Schechter function and
the dashed line is the Schechter function fit of
Ebeling et al. \protect\markcite{e97}(1997) 
to their low redshift RASS cluster sample.
\label{fig:rosat-xlf}}
\end{figure}

\begin{figure}[t]
\plotfiddle{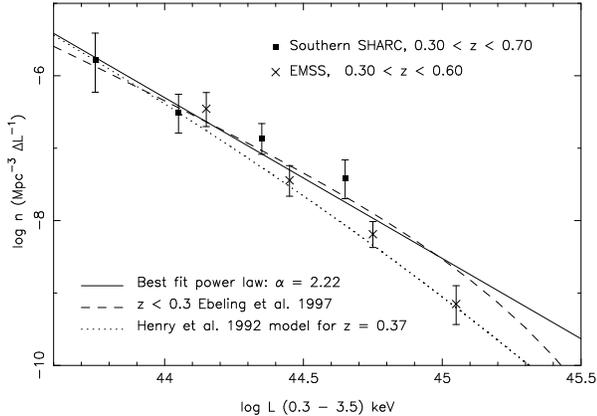}{2in}{-90}{32}{32}{-120}{180}
\caption[fig4.eps]{
The solid points show the Southern SHARC 
XLF in the 0.3 to 3.5 keV pass band and
the solid line is our best power-law fit.
The crosses indicate the high redshift XLF
from Henry et al. \protect\markcite{h92}(1992)
with the dotted line showing their best fit model
($n = -2.10$, $k_0 = 0.029 h$ Mpc$^{-1}$)
evaluated at a redshift of 0.37, the median
of the combined cluster sample.
The dashed line is the Schechter function fit of
Ebeling et al. \protect\markcite{e97}(1997).
\label{fig:emss-xlf}}
\end{figure}

In Figures \ref{fig:rosat-xlf} and \ref{fig:emss-xlf} we show the 
non-parametric XLF, for the redshift shell $z = 0.3 - 0.7$,
calculated using equations (\ref{eqn:vmax}) and (\ref{eqn:xlf-binned})
for both the 0.5 - 2.0 keV and 0.3 - 3.5 keV pass bands.
Table \ref{tbl:xlf-binned} lists the data points shown in these figures.

We fitted two parametric forms to the data: 
a power law (e.g. Henry et al. \markcite{h92}1992),
\begin{equation}
\label{eqn:xlf-power-law}
\phi(L) = K  L^{-\alpha}, 
\end{equation}
and a Schechter function (e.g. Ebeling et al. \markcite{e97}1997),
\begin{equation}
\label{eqn:xlf-schechter}
\phi(L) = K_s  \exp( -L / L_* ) \, L^{-\alpha_s}.
\end{equation}
As we are not sensitive to the characteristic luminosity, $L_*$,
of the Schechter function, we used the best fit values from
Ebeling et al. \markcite{e97}(1997) --- 
$5.70 \times 10^{44}$ \lumin\ (0.5-2.0 keV) and
$10.7 \times 10^{44}$ \lumin\ (0.3-3.5 keV).
The likelihood for a given $\phi(L)$ is independent of the 
normalisation ($K$ or $K_s$), 
so we fixed this by setting the expected number of 
clusters equal to the number in our sample.
This procedure gives some weight to luminosities where no clusters
were detected, producing a steeper slope than the non-parametric XLF 
would suggest (Henry et al. \markcite{h92}1992).
Errors on $\alpha$ and $\alpha_s$ were calculated by integrating the normalised
likelihood distributions, which are well approximated by a Gaussian,
out to an enclosed area of 68.3\%
and those for the normalisations were found by allowing $\alpha$ to take
its one-sigma values.
The results are listed in Table \ref{tbl:xlf-fit}
and shown graphically in
Figures \ref{fig:rosat-xlf} and \ref{fig:emss-xlf}.

\section{Discussion}
\label{discussion}

From Figure \ref{fig:rosat-xlf} it is qualitatively obvious that,
for $L \sim 10^{44}$ erg s$^{-1}$, our $z = 0.3 - 0.7$ cluster XLF is 
consistent with that of the low redshift cluster XLF.
Our best fit faint end slope for the Schechter function,
$\alpha_s = \srabest \pm \srasigma$, agrees with that of the
low redshift cluster sample of Ebeling et al. (1997),
who find $\alpha_s = 1.85 \pm 0.09$.
The power-law slope in the EMSS pass band ($\alpha = \aebest \pm \aesigma$) 
is also consistent with the slope of the low redshift shell of
Henry et al. \markcite{h92}(1992) and the re-worked EMSS sample of 
Nichol et al. \markcite{n97}(1997), for which $\alpha = 2.19 \pm 0.21$ and 
$\alpha = 2.60 \pm 0.37$ respectively. The lack of luminosity evolution 
agrees with our recent analysis of the
redshift distribution of the Southern SHARC clusters
(Collins et al. \markcite{c97}1997) and is inconsistent with the 
strong evolution claimed by Castander et al. \markcite{RIXOS}(1995). 
Our results strengthen, and extend to higher redshift, recent claims that 
cluster properties remain the same out to $z\simeq 0.3$
(e.g. Ebeling et al. \markcite{e97}1997;
Mushotzky \& Scharf \markcite{ms97}1997). 

The only other published XLF for X-ray selected clusters covering our
redshift range is that of Henry et al. \markcite{h92}(1992) from 21 EMSS
clusters between $z=0.3-0.6$, with a median redshift of 0.33.
They find a much steeper power-law slope of
$\alpha=3.27\pm0.29$, which they claim is the result of negative
evolution.
Our XLF is consistent with the non-parametric XLF of 
Henry et al. \markcite{h92}(1992)
for luminosities $\lesssim 5 \times 10^{44}$ \lumin\ (0.3 - 3.5 keV),
as shown in Figure~\ref{fig:emss-xlf}, hence any evolution of the
cluster population is restricted to luminosities $\gtrsim 3 \times 10^{44}$ \lumin.
Using the Ebeling et al. \markcite{e97}(1997) XLF,
and assuming 100\% detection efficiency,
we find that the surface density of clusters with $L > 3 \times 10^{44}$
\lumin, in the redshift range 0.3 to 0.7, 
is 0.083 deg$^{-2}$ 
and the expected number of such clusters in the 
Southern SHARC survey is only $\sim 1.5$.
Therefore the lack of high luminosity clusters in our sample is not unexpected.
In Figure \ref{fig:emss-xlf}, we plot equation (6) of 
Henry et al. \markcite{h92}(1992) using their best-fit values of
$n = -2.10$ and $k_0 = 0.029 h$ Mpc$^{-1}$, for $z = 0.37$,
corresponding to the median redshift of the joint cluster sample. 
This analytical model is based on the Press-Schechter formalism and is 
shown here simply to provide a physical basis for extrapolating the EMSS 
data and comparing the two samples. 
In the context of general cluster formation 
theories, our result is broadly 
consistent with the predictions of entropy-based models 
or a low value of the density parameter
(e.g. Kaiser \markcite{k91}1991; Evrard \& Henry \markcite{eh91}1991; 
Henry et al. \markcite{h92}1992; Oukbir \& Blanchard \markcite{oukbir-97}1997; 
Bower \markcite{bower-97}1997; Mathiesen \& Evrard \markcite{mathiesen-97}1997).

\section{Conclusion}
\label{conclusion}

We have used the Southern SHARC survey to create the first
$z > 0.3$ XLF derived from ROSAT detected clusters of galaxies.
Comparison with the low redshift cluster XLF of both ROSAT 
(Ebeling et al. \markcite{e97}1997) and EMSS 
(Henry et al. \markcite{h92}1992) clusters
shows that there is no evolution in the X-ray luminosities of 
$L \sim 10^{44}$ erg s$^{-1}$ clusters at a median depth of $z=0.44$.
This is consistent with our analysis of the
redshift distribution of this cluster sample 
(Collins et al. \markcite{c97}1997)
and adds further weight to the body of evidence for no evolution 
in the cluster population.

\acknowledgements

This research has made use of data obtained from the Leicester Database
and Archive Service at the Department of Physics and Astronomy,
Leicester University, UK.
DJB acknowledges PPARC for a Postgraduate Studentship and PDRA, 
CAC acknowledge PPARC for an Advanced Fellowship,
AKR acknowledges support from NASA ADP grant NAG5-2432, and
BH acknowledges support by the Center for Astrophysical Research in
Antarctica (NSF OPP 89-20223).




\begin{deluxetable}{cccc}
\tablewidth{0pt}
\tablecaption{The non-parametric XLF for the Southern SHARC survey.\label{tbl:xlf-binned}}
\tablehead{
\colhead{pass band} &
\colhead{$\log L$\tablenotemark{a}} & 
\colhead{$\log n(L)$\tablenotemark{b}} \\
\colhead{(keV)} &
\colhead{} & 
\colhead{} 
}
\startdata
$0.5 - 2.0$ & 43.50 & $-5.52$ $(+0.37,-0.45)$ \nl
$0.5 - 2.0$ & 43.80 & $-6.26$ $(+0.25,-0.28)$ \nl
$0.5 - 2.0$ & 44.10 & $-6.61$ $(+0.20,-0.22)$ \nl
$0.5 - 2.0$ & 44.40 & $-7.17$ $(+0.25,-0.28)$ \nl
\tableline	    	      
$0.3 - 3.5$ & 43.75 & $-5.78$ $(+0.37,-0.45)$ \nl
$0.3 - 3.5$ & 44.05 & $-6.51$ $(+0.25,-0.28)$ \nl
$0.3 - 3.5$ & 44.35 & $-6.86$ $(+0.20,-0.22)$ \nl
$0.3 - 3.5$ & 44.65 & $-7.42$ $(+0.25,-0.28)$ \nl
\enddata
\tablenotetext{a}{$L$ has units of erg s$^{-1}$}
\tablenotetext{b}{$n(L)$ has units of Mpc$^{-3}$ $\Delta L^{-1}$}
\tablecomments{
Luminosity bins have a constant width
and were chosen so that each bin contained at least two clusters.
Poisson errors (one-sigma) are from Gehrels \protect\markcite{gehrels-86}(1986).
}
\end{deluxetable}

\begin{deluxetable}{ccccc}
\tablewidth{0pt}
\tablecaption{The parametric XLF for the Southern SHARC survey.\label{tbl:xlf-fit}}
\tablehead{
\colhead{pass band} & 
\colhead{$\alpha$}        & 
\colhead{$K$\tablenotemark{a}} &
\colhead{$\alpha_s$}        & 
\colhead{$K_s$\tablenotemark{a}} \\
\colhead{(keV)} & 
\colhead{}      & 
\colhead{}      &
\colhead{}      & 
\colhead{}
}
\startdata
0.5 - 2.0  & $\arbest \pm \arsigma$ & $\krbest \pm \krsigma$ & $\srabest \pm \srasigma$ & $\srkbest \pm \srksigma$ \nl
0.3 - 3.5  & $\aebest \pm \aesigma$ & $\kebest \pm \kesigma$ & $\seabest \pm \seasigma$ & $\sekbest \pm \seksigma$ \nl
\enddata
\tablenotetext{a}{Units of \kunits}
\end{deluxetable}


\begin{references}
\reference{bower-97}
Bower, R. G. 1997, \mnras, in press, astro-ph/9701014
\reference{RIXOS}
Castander, F. J., et al. 1995, \nat, 377, 39
\reference{c97}
Collins, C. A., Burke, D. J., Romer, A. K., Sharples, R. M.,
\& Nichol, R. C. 1997, \apjl, 479, L117
\reference{e97}
Ebeling, H., Edge, A. C., Fabian, A. C., Allen, S. W., 
Crawford, C. S., \& B\"ohringer, H. 1997, \apjl, 479, L101
\reference{eh91}
Evrard, A. E., \& Henry, J. P. 1991, \apj, 383, 95
\reference{gehrels-86}
Gehrels, A. 1986, \apj, 303, 336
\reference{gl-94}
Gioia, I. M., \& Luppino, G. A. 1994, \apjs, 94, 583
\reference{h92}
Henry, J. P., Gioia, I. M., Maccacaro, T., Morris, S. L., Stocke, J. T., 
Wolter, A. 1992, \apj, 386, 408
\reference{jf84}
Jones, C. \& Forman, W. 1984, \apj, 276, 38
\reference{k86}
Kaiser, N. 1986, \mnras, 222, 323
\reference{k91}
Kaiser, N. 1991, \apj, 383, 104
\reference{mathiesen-97}
Mathiesen, B., \& Evrard, A. E. 1997, \mnras, submitted, astro-ph/9703176
\reference{ms97}
Mushotzky, R. F., \& Scharf, C. A. 1997, \apjl, 482, L13
\reference{n97}
Nichol, R. C., Holden, B. P., Romer, A. K., Ulmer, M. P.,
Burke, D. J., \& Collins, C. A. 1997, \apj, 481, 644
\reference{oukbir-97}
Oukbir, J., \& Blanchard, A. 1997, \aap, 317, 1
\reference{sarazin}
Sarazin, C. L. 1988, X-ray emission from clusters of galaxies
(Cambridge: Cambridge University Press)
\reference{ws93}
Wang, Q., \& Stocke, J. T. 1993, \apj, 408, 71
\end{references}
\end{document}